# A Neural Network Based Framework for Archetypical Sound Synthesis

Eric Guizzo
Conservatorio C.Pollini
of Padua
eric.guizzo@hotmail.it

Alberto Novello
Conservatorio C.Pollini
of Padua
albynovello@gmail.com

## ABSTRACT

This paper describes a preliminary approach to algorithmically reproduce the archetypical structure adopted by humans to classify sounds. In particular, we propose an approach to predict the human perceived *chaos/order* level in a sound and synthesize new timbres that present the desired amount of this feature. We adopted a Neural Network based method, in order to exploit its inner predisposition to model perceptive and abstract features. We finally discuss the obtained accuracy and possible implications in creative contexts.

•Applied computing→Performing arts •Information systems→ Multimedia and multimodal retrieval

## 1. INTRODUCTION

Sound perception is able to recall metaphoric and synesthetic sensations in human mind. Humans often adopt these sensations to describe and categorize audio-related events. It is common for instance, to describe the sound produced by an old closing door or a broken celery as *crackly*. These timbres are different, but they share certain features that make humans associate them with the same archetype. The latter is intended as a semantic label that qualitatively represents the formal appearance of a sound category (in this case, the *crackle*). Human imagination processes are related to archetypical structures [1]. In this paper we describe an algorithmic model of sound archetypes to computationally represent and manipulate audio information in a human-like fashion, directly reflecting our inner process of conceiving sounds.

## 2. BACKGROUND

In *Traité des Objets Musicaux* [2]*,* Pierre Shaeffer identified an absolute paradigm through which univocally classify sound events by specifying a selected set of perceptive features evoked by the sounds. This criterion allowed the identification of timbre classes, characterized by specific qualities. An important corollary of this theory is the concept of *Temporal Semiotic Unit* (TSU) [3]. The TSU is based on the evolution from the notion of *Sound Object* to the idea of a *Semiotic Sound Object*. This implicates a separation from Shaeffer's pure gestalt-oriented view, which considered the timbre as an entity completely isolated from its context, relying on the conception of a "limited listening" that ignores any "causal or associative meaning" of the sound object [3]. Conversely, TSUs take in consideration the *semantic value* of audio material organized as temporal concatenation of sound events. This concept can be applied both to temporal succession of sound objects (i.e. musical figures) and to the evolution (through time) of the timbre structure within a single sound object. The latter aspect is of particular interest for this research because it allows to identify TSUs as *sound archetypes*. It has emerged that these perceptive characters are strictly dependent on the background and the experience of an individual [3]. This is also empirically evident, in fact, for instance, the perception of a *happy* sound could mean something different for diverse individuals. Then, it is fundamental to consider the *ambiguity level* of a sound archetype. Accordingly, the higher is the ambiguity, the higher is the perception subjectivity of an archetype. Therefore, from a theoretical point of view, this research aims to algorithmically model TSUs, taking into account their intrinsic ambiguous character.

Two sub-categories of *Music Information Retrieval* are of strong interest for this project: a*utomatic signal classification,* based on the semantic content of the sound files*,* and a*utomatic data synthesis,* involving the generation of sounds that match specific signal-level and perceptive features.

Automatic signal classification approaches based on *Neural Networks* (NN) demonstrated a solid accuracy, outperforming traditional algorithms that rely on *handcrafted feature extraction,* especially for large-data tasks [4]. NNs are capable of analyzing complex information (as sounds, images and videos) in a similar way as human brain does and perform complex operations among gathered data, such as finding similarities and infer data categories. Then, NNs can be *trained* to confer them a particular experience to affine the performance for a specific task (for example song genre recognition) [5]. In a similar manner, a human learns the correct movements required to ride a bicycle by trial and error. By exploiting this human-like behavior, they are efficient in computing fuzzy or undefined features, which are often involved in applications related to audio-visual perception [6]. The NN approach can overcome the necessity of precisely expressing complex mathematical structures that represent the features to be extracted, which can be problematic in the case of human perception [5]. On the other hand, this approach presents two fundamental drawbacks: first, NNs must be trained with a large amount of data in order to reach a reasonable experience, and then precision. Second, a consistent computing power is often required. Various studies point out that *Convolutional Neural Networks* (CNN) and *Recurrent Neural Networks* (RNN) are the most suitable for audio applications. In particular, it has been proved that CNNs perform better in tracking static features, whereas RNNs are more convenient for time-related dependencies [6][7]. Then, RNNs could seem the obvious choice for audio-related applications, considering the intrinsic sequential nature of sound information. Nevertheless, *Z*hang et al. [8] demonstrated that an entirely CNN-based architecture can perform with comparable accuracy on audio classification tasks (speech recognition), providing a significantly higher computing efficiency with respect to RNNs. A combination of the two methods has been ascertained to merge the benefits of both strategies, notwithstanding the high computing requirements. *Choi et al.* [7], for instance, have successfully adopted this approach.

Feature-matching audio synthesis applications have adopted different approaches, using statistical models based on *Markov Chains* and *Neural Networks*. Markov Chains are appropriate to generate sequential data, emulating the behavior of a given experience. This makes the latter an ideal method for musical composition tasks (sequences of notes). An illustrious example of this application has been undertaken by Iannis Xenakis in "Analogique" [9]. In addition to this, markovian processes have been extensively used to perform timbre-level audio synthesis, considering that the set of parameters required by any sound synthesis algorithm could be intended as a (non temporal) sequence. *Hidden Markov Model Vocoders (HMMV)* [10] represent a common utilization of this method, although they have been mostly applied to perform automatic text to speech synthesis. NN-based synthesis approach can be adopted for the generation of the features to be transformed in audio, whereas other specific algorithms synthesize the final waveform. *Zen et al.* [11] demonstrated that this technique could surpass the accuracy of previous state of art methods, which were based on markovian processes. On the other hand, NNs are proved to be enough powerful to directly compute the output waveform sample by sample. Google, with Wavenet [12], has proposed an important example of this technique. Nevertheless, notwithstanding the average better accuracy of NNs, the Markov Chain approach permits to spare considerable computing resources, especially for simple models [13].

Several techniques have been developed to synthesize sounds matching specific *perceptive* features. NN-based techniques are intrinsically suited for this task. In fact, NNs can be viewed as algorithmic structures that follow rules analogous to the gestalt laws of grouping [14], reproducing the inner predisposition of human mind to extract patterns from perceptive stimuli and to adopt them (instead of a raw representation of the stimuli) to categorize and recall experiences. One notable example has been undertaken by G*ounaropouls et al.* [15], presenting a NN-based methodology to synthesize sounds starting from given perceptive adjectives.

Feature-matching audio synthesis is still a relatively new domain of research, with the majority of cases focusing on speech synthesis. The above-mentioned researches (and many others) suggest that archetypical sound synthesis is not only possible, but can lead to satisfactory results, in particular adopting NN and Markov Chain based approaches. However, these methods have been rarely applied for artistic purposes. From a practical point of view, the objective of this research is to produce a working and usable framework to perform features-matching audio synthesis based on human perception. The goal is to obtain an optimized environment capable of being employed in real-time on a common laptop computer. In the present article, we consider to model only one sound archetype: the amount of human-perceived *chaos/order* in audio information. We selected this characteristic in particular for its intrinsic fuzziness and non-specific connotation (compared to other perceptive features, such as *metallic/wooden*). Therefore, in order to achieve a faster and deeper comprehension on what will follow, we encourage the reader to think about what determines a sound to be chaotic and which characteristics should have to be ordered, basing on his personal perception and experience.

## 3. METHOD

The workflow we followed is divided in 4 consecutive stages: dataset creation, development of the analysis algorithm, development of the re-synthesis algorithm, development of the user interface. This has been fixed a prior, considering to adopt a NN-based approach, in order to exploit the predisposition of this method to model perceptive and abstract features. Furthermore, our particular approach is based on a critical consideration: perception-related phenomena can be studied following a Helmoltzian or a gestalt-oriented view. The first focuses on the identification, quantification and interpretation of neurophysiological processes derived from perceptive stimuli. Instead, the latter is oriented to the interpretation of the abstract sensations produced by the same stimuli, which are intrinsically non-measurable entities and can be analyzed only through human descriptions [16]. We consider the latter a more efficient way to achieve our task since sound imagination concerns abstract emotions that are non-exactly measurable, culture-dependent and subjective.

The homogeneity of the training dataset is a crucial aspect for the effectiveness of a NN. In fact, it has been proved that non-homogeneous datasets (containing largely different amounts of data-points for each classification label) tend to produce unbalanced and inaccurate outcomes [17]. For this reason we have paid great attention to collect a dataset the more balanced as possible. Initially, we downloaded 100 sound samples from the Freesound database [18]. The files have been randomly chosen, in order to minimize any bias that could derive from our personal influence in the selection. Successively, the samples have been processed through a granular synthesis algorithm in order to extract textures from the sounds. This procedure permitted to obtain multiple different timbres from a single sound file. A set of 1000 3-seconds textures has been recorded, trying to obtain a balanced dataset, with en equal amount of chaotic and ordered sounds. For this expedient it has been necessary to impose decisions based on our personal perception. This certainly included bias, although we retained this aspect secondary to the homogeneity of the dataset. Four different subjects have rated each sound sample on the perceived *order* level, in the form of a discrete mark (from 0 to 10). These human classifications are necessary to implement a *supervised* learning process, since they represent the correct predictions expected from the classification algorithm [5]. After this classification, certain data-points have been discarded in order to maintain an equal quantity of timbres for each label. The files to be eliminated have been chosen according to the p-value obtained by performing a normality test on single sounds' classifications. This eliminated the data-points with the higher ambiguity level, keeping the ones for which the sound *order* level can be considered equally perceived by distinct persons. We adopted the Shapiro-Wilk test for this purpose, since it has been proved to be effective for low amount of samples [19], as this case requires. Finally, the arithmetic mean of the classifications has been computed to obtain data-points labeled with a single value. After this stage, the dataset has been processed through an augmentation algorithm. This operation is aimed at reducing overfitting problems in the NN training, as pointed out by *Wang et al.* [20]. The augmentation technique we adopted involves a series of elaborations that generate slightly different versions of every sound, maintaining the amount of *order* present in the original files (this property has been empirically verified through informal surveys). These elaborations have been individually applied to every sound:
-Convolution between high-energy spectral areas
-Random filtering
-Random time stretching with pitch shift
-Convolution with random room impulse responses
The augmentation process extended the dataset from 425 to 4675 data-points.

The analysis block consists of an algorithm that aims at predicting the human-perceived order level of an audio signal. For this task, we employed a Convolutional Neural Network design, to reach a reasonable accuracy with contained computing costs, as explained above. The model has been trained with sounds' spectra, adopting the human classifications as target labels. The training has been performed for 50 epochs with a batch size of 10 data-points and using *Categorical Crossentropy* as loss function. The implemented layer architecture follows the scheme of Figure 1:

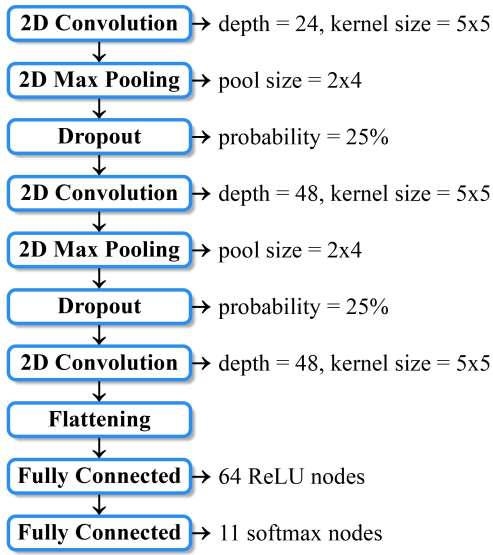

**Figure 1: CNN layer design**

This layout was inspired by the work of *Salomon et al.* [21], although substantial modifications have been applied. In particular, instead of training the network with MFCCs, a simpler STFT provided more accurate predictions for our particular purpose. We adopted 80% of the dataset for the CNN training, the 10% as validation set, and the remaining 10% as test set to assess the final model accuracy on unobserved data. This architecture presents a 98,5% classification accuracy for the training and 96,7% for the test set, which could be considered a satisfying result at this stage. Finally, the network has been arranged for a pseudo real-time utilization (around 100 milliseconds latency), in order to be able to continuously classify audio streams coming, for instance, from microphones.

The synthesis section of the environment aims to generate on demand sound textures that present a desired amount of *order*. This has been obtained adopting the same granular synthesis algorithm employed for the dataset creation, but generating its parameters through a Markovian process (instead of randomly). This workflow has been selected in order to consume the lowest amount of computing resources and maintain the possibility of operating the framework on a common laptop. In fact, we empirically denoted that an NN-based approach for sound synthesis (such as Google's *Wavenet*) is exceedingly computing expensive. To build the synthesis model, first a dataset containing 2000 granular synthesis parameters-sets has been collected. For this purpose, random sets have been created and labelled with their relative *order* level, by classifying the correspondent generated sound with the above-described CNN. Again, the dataset has been built collecting the same amount of of data-points for each label (0-10). This dataset contains then the transition matrixes that a simple markov chain adopts to generate sequences of synthesis parameters that produce sound textures with a desired amount of human-perceived *order*. The accuracy of this process has been validated through a formal survey, comparing the requested *order* levels with human classifications about the synthesized sounds. The opinions we collected indicate an average re-synthesis accuracy of 95,5%. Nevertheless, it emerged that a request of chaotic synthesis parameters always produces a chaotic texture. On the contrary, a request of ordered synthesis parameters sometimes produces "softly disordered" sounds.

At the end, the classification and synthesis algorithms have been connected in a single framework with a simple user interface. This provides the possibility of continuously classifying a signal coming from a microphone (containing, for example, the self output of the system diffused by a speaker mixed with sounds generated by people present in the room) and producing sounds with a desired amount of *order*. These processes can also be automated to obtain constantly changing textures. Furthermore, it is possible to make the sounds to be generated directly dependent from the CNN predictions, for example producing textures that present opposite *order* levels. This creates an environment that constantly "listens to itself" and tries to balance the amount of chaos and order in the room, reacting also to noises emitted by the listeners. In addition to these capabilities, two supplementary audio processing algorithms have been added: a reverb and a random amplitude amplifier. We empirically noticed that applying a plate-like reverberation to a texture always makes the CNN predictions more oriented towards order. Conversely, imposing random amplitude modulations shifts the predictions towards chaos. Then, through these two simple controls, it is possible to *bias* the CNN, moving the predictions towards a certain direction. Therefore, the user interface brings control over the order level of a sound to be generated, automation modalities and audio processing. The usable framework can be summarized with the following block diagram:

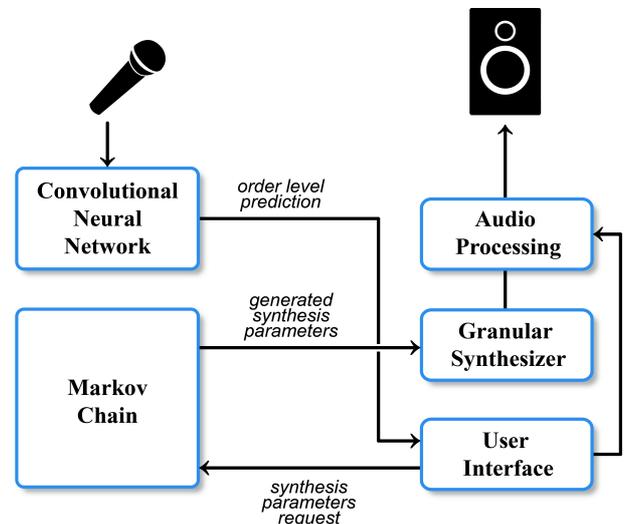

**Figure 2: Environment block diagram**

This scheme refers to one single instance of the environment.

## 4. CONCLUSIONS

We described a real-time algorithmic environment for archetypical sound analysis and re-synthesis. Currently only one feature has been modeled: the human-perceived *order* in sound. The developed framework is capable of predicting the amount of this feature present in audio signals, as well as synthesizing sounds presenting a desired amount of this perceptive character. The analysis algorithm reached 96,7% accuracy for the test dataset, obtained evaluating the classification model on data unobserved in the training process. The average re-synthesis accuracy is equal to 95,5%, obtained comparing the requested *order* level of sounds generated by the algorithm with human judgments. The perceptive opinions we collected so far indicate that the re-synthesis produces textures "slightly more chaotic" than requested. In order to improve the classification accuracy we intend to investigate different CNN architectures, as well as a RNN-based implementation. Furthermore, once obtained a definitive measure of the re-synthesis accuracy, an RNN-based synthesis approach will be considered as well.

Despite the substantial restrictions of the actual environment, it suggests new expressive approaches to music production, improvisation and interaction design. In particular, it allows the artist to create sounds based on replicating human perception. Moreover, it permits to classify any audio signal according to the same criterion and adopt the predictions to control any parameter of a synthesis or processing algorithm.

For future implementations, we intend to extend the palette of perceptive features. The goal is to model a significant amount of low-ambiguity-level characteristics (that can be univocally conceived by different individuals). In addition to this, we intend to improve the algorithm in order to model archetypes basing on sharply restricted training datasets. This implies, in particular, to improve our dataset augmentation algorithm and make it reliable for any kind of archetype (instead of the only *sound order*). This would provide the user the possibility to construct his own models with little effort and then to represent his subjective archetypes, reflecting his personal manner to think sounds.